\documentclass[aps,prl,groupedaddress,float, superscriptaddress, reprint]{revtex4-2}

\usepackage[english] {babel}
\usepackage{amsfonts,graphicx,soul,natbib,mathrsfs}

\usepackage[colorlinks=true]{hyperref}
\usepackage{appendix}
\usepackage{epstopdf}
\usepackage{soul}
\usepackage{physics}
\usepackage{graphics}
\usepackage{comment}
\usepackage{bm}
\usepackage{amsfonts,amssymb,latexsym,amsmath,enumerate,amsthm}
\usepackage{subcaption}

\newcommand{\be}{\begin{equation}}
 \newcommand{\ee}{\end{equation}}
\newcommand{\bear}{\be\begin{array}}
\newcommand{\bea}{\begin{eqnarray}}
\newcommand{\eea}{\end{eqnarray}}

\newcommand{\bp}{{\bf p}}
\newcommand{\br}{{\bf r}}

\newcommand{\la}{\langle}
\newcommand{\ra}{\rangle}

\usepackage[dvipsnames]{xcolor}

\newcommand{\dst}{\displaystyle}
\newcommand{\fr}[2]{\frac{{\dst #1}}{{\dst #2}}}

\begin{document}

\title{Angular momentum dynamics \\ of vortex particles in accelerators}

\author{D.\,Karlovets}
\affiliation{School of Physics and Engineering,
ITMO University, 197101 St. Petersburg, Russia}%

\affiliation{Petersburg Nuclear Physics Institute of NRC “Kurchatov Institute”, Gatchina 188300, Russia}
\author{D.\,Grosman}
\affiliation{School of Physics and Engineering,
ITMO University, 197101 St. Petersburg, Russia}
\author{I.\,Pavlov}
\affiliation{School of Physics and Engineering,
ITMO University, 197101 St. Petersburg, Russia}

\affiliation{Petersburg Nuclear Physics Institute of NRC “Kurchatov Institute”, Gatchina 188300, Russia}

\date{\today}

\begin{abstract}

While conventional experiments typically employ plane-wave states of particles with definite momenta, vortex states represent cylindrical waves carrying an orbital angular momentum (OAM) projection along the propagation direction. This projection can be arbitrarily large, granting charged particles magnetic moments orders of magnitude greater than those of plane-wave states. Consequently, vortex beams could complement or replace spin-polarized beams in high-energy collisions, accessing observables beyond the reach of conventional experiments. We investigate the radiative and non-radiative OAM dynamics for relativistic vortex particles in accelerators. Our results show that the timescale for OAM loss via photon emission significantly exceeds typical acceleration times. Non-radiative OAM dynamics is governed by precession at a frequency distinct from that of spin. Similar to spin tunes, this induces resonances that can disrupt OAM at much lower energies than for spin-polarized beams. Thus, we propose using linacs for acceleration of the vortex beams, while Siberian snakes can be adapted for OAM manipulations.
\end{abstract}

\maketitle

\textit{Introduction---} Spin-polarized electrons, protons, ions, and other particles are needed for existing and future accelerators \cite{shiltsev2021modern, xue2025compact, mane2005siberian, mane2005spin}, including plasma-based ones \cite{buscher2020generation, nie2021situ, nie2022highly, sun2022production}, and the means to control and preserve polarization while increasing the energy are crucially important for them. Charged particles with a quantized projection of orbital angular momentum (OAM) on a propagation axis $\ell = 0,\pm \hbar, \pm 2\hbar,...$ \cite{Bliokh2011, Bliokh2017}, dubbed vortex particles, have a magnetic moment that grows with the OAM, $|{\bm \mu}| \propto |\ell|$ \cite{Bliokh2011, Bliokh2017}, and it is tempting to think that they can also be used in high-energy physics along with the spin-polarized beams or even instead of them \cite{Ivanov2012, Ivanov2016, Bliokh2017, karlovets2017scattering, KPO2017, silenko2017manipulating, silenko2018relativistic, silenko2019siberian, karlovets2020effects, Ivanov2020, karlovets2021vortex, ivanov2022promises, zou2022production, Huizhou}. It has been predicted that by colliding spin-unpolarized vortex states, one can generate spin-{\it polarized} particles and control their polarization \cite{ivanov2020doing}.

Beyond the fundamental interest in accelerating exotic quantum states to enable fixed-target and collision experiments inaccessible with ordinary beams, both charged and neutral vortex particles -- the latter being twisted neutrons \cite{Pushin2016, Pushin2022} and atoms \cite{Luski2021} -- can bring novel insights into strong interactions, especially at low energies \cite{Ivanov2012, Ivanov2016, ivanov2022promises, Bliokh2017, karlovets2017scattering, karlovets2020effects, DK2021}, quantum measurements and entanglement at high energies \cite{karlovets2023shifting}, quantum interference and coherence phenomena, usually inaccessible in particle-physics experiments \cite{kazinski2023inclusive, akimov2025plasmon, karnieli2021coherence, karlovets2025attosecond, ruimy2025free, Pushin2016, Pushin2022}, electron microscopy \cite{uchida2010generation, beche2014magnetic, verbeeck2010production, Bliokh2017, Schattschneider2013, verbeeck2011atomic}, neutron interferometry and holography \cite{Sarenac2016, Pushin2016, Pushin2022}, and so forth. However, so far no ultrarelativistic vortex particles have been obtained at an accelerator facility.

Here we argue that vortex particles can be accelerated in both linear and circular accelerators and their OAM can be manipulated by using the same techniques as spin, for instance, with Siberian snakes \cite{mane2005siberian, mane2005spin}. It is the magnetic - {\it not mechanical} - moment that interacts with electromagnetic fields, and its temporal dynamics defines that of spin and the OAM. The latter is governed by {\it (i)} precession of the OAM magnetic moment, and {\it (ii)} by the radiative OAM-depolarization due to the loss of the OAM during emission of twisted photons. Using quantum electrodynamics (QED), we calculate an effective time of the OAM loss in magnetic field, finding it many orders of magnitude larger than an acceleration time. These findings enable the acceleration of vortex beams, facilitating their use in fixed-target and collider experiments.

The non-radiative OAM dynamics differs drastically from that of spin. By generalizing the Bargmann–Michel–Telegdi (BMT) equation \cite{bargmann1959precession} to include intrinsic OAM, we predict OAM resonances analogous to spin tunes \cite{froissart1960depolarisation, Derbenev1978, mane2005siberian, mane2005spin}. These can induce non-radiative OAM-depolarization of the vortex beams in circular accelerators. For electrons, OAM resonances emerge at $\varepsilon\approx 3$ MeV with a $\Delta \varepsilon \approx 1$ MeV  interval, whereas the first spin resonance due to the anomalous magnetic moment (AMM) occurs only at $\varepsilon \approx 440$ MeV. This discrepancy arises because the OAM contribution, coupled with Thomas precession, is significantly stronger than the AMM effect, and it can easily be probed even with small machines like microtrons if vortex electrons were accelerated to the MeV range. These dynamics are directly relevant to current projects, including the 200-MeV vortex electron source at JINR (aiming for 400 MeV by 2027) \cite{JINR}, the vortex ion source in Huizhou \cite{Huizhou}, etc.

The use of Siberian snakes can therefore be of higher importance for vortex beams in synchrotrons and storage rings than for spin-polarized beams, because without them the OAM-polarization can quickly degrade. Linacs do not face these challenges, which is why they can be used to accelerate vortex particles, thereby enabling a plethora of completely new types of experiments in particle physics, unthinkable with ordinary beams \cite{ivanov2022promises}. The system of units $\hbar=c=1$ is used.


\textit{Quantum dynamics of the particle magnetic moment---} Let an operator of kinetic momentum be $\hat{\bp}$ and its canonical counterpart $\hat{\bp}^{(c)} = \hat{\bp} + e{\bm A} = -i{\bm \nabla}$ where $e$ is the particle charge whose sign is arbitrary, and ${\bm A}$ is a potential of the electromagnetic field. There are two operators of OAM, $\hat{\bm L}^{(c)} = \br\times \hat{\bp}^{(c)}$ and $\hat{\bm L} = \br\times \hat{\bp}$, the latter being gauge-invariant while the former being not, and $\br$ is a non-relativistic position operator. Let $\br_0$ be a vector from the origin to the axis relative to which the OAM is measured. We rewrite 
\bea
\hat{\bm L}^{(c)} = \br\times \hat{\bp}^{(c)} = \underbrace{\br_0\times \hat{\bp}^{(c)}}_{\hat{\bm L}^{(c)}_{\text{ext}}} + \underbrace{\left(\br-\br_0\right)\times \hat{\bp}^{(c)}}_{\hat{\bm L}^{(c)}_{\text{int}}}.
\eea
Here $\hat{\bm L}^{(c)}_{\text{ext}}$ is {\it extrinsic} because it can be put to zero by shifting the origin, $\br_0\to 0$, whereas $\hat{\bm L}^{(c)}_{\text{int}}=\left(\br-\br_0\right)\times \hat{\bp}^{(c)}$ is {\it intrinsic} because $\br-\br_0$ is invariant with respect to the shift. In what follows, we put $\br_0=0$, so the OAM is intrinsic everywhere below.

We begin in the co-moving (rest) frame in which a mean momentum of the packet vanishes, $\la \bp \ra=0$. 
A non-relativistic operator of the spin magnetic moment is \cite{Landau1981Quantum}
\bea
\displaystyle & \hat{{\bm \mu}}_s = 2\mu_s \hat{{\bm s}} = -g\mu_B \hat{{\bm s}},\ \hat{\bm s} = {\bm \sigma}/2,\cr
\displaystyle &  g=-2\mu_s/\mu_B=\text{inv},\ \mu_B = \fr{|e|}{2m}\left(\equiv\fr{|e|\hbar}{2mc}\right),
\eea
where $m$ is the particle mass, ${\bm \sigma}$ are the Pauli matrices, and $\mu_B$ is the Bohr magneton. When an orbital part is added, an operator of the magnetic moment interacting with a field is given by Eq.(47.9) in \cite{BLP},
\bea
\hat{{\bm \mu}}^{(c)} = -\mu_B \left(\hat{{\bm L}}^{(c)} + g\hat{{\bm s}}\right),
\label{munosoi}
\eea
where the OAM is canonical and hence $\hat{{\bm \mu}}^{(c)}$ is {\it not} gauge-invariant. Its gauge-invariant kinetic counterpart is  
\bea
\hat{{\bm \mu}} = -\mu_B \left(\hat{{\bm L}} + g\hat{{\bm s}}\right).
\label{mukin}
\eea

Next, we neglect the spin-orbital interaction (SOI) but keep the AMM, so $g\ne 2$. The mean magnetic moment of a free relativistic fermion with a phase vortex is \cite{Bliokh2011, Bliokh2017, PhysRevA.98.012137}
\bea
\displaystyle & \la \hat{{\bm \mu}}\ra = -\mu_B \left(\ell \hat{{\bm z}} + g\la\hat{{\bm s}}\ra + \Delta\right),\ \hat{{\bm z}} = (0,0,1),
\label{munosoi2}
\eea
where 
$\Delta$ is due to the SOI. 
Although the SOI effects were shown to be enhanced for $|\ell| \gg 1$ \cite{PhysRevA.98.012137}, a realistic estimate of them is mush smaller than the AMM:
\bea
\Delta \sim |\ell|\fr{\lambda_c^2}{\sigma_{\perp}^2} < 10^{-5},\quad
a = \fr{g-2}{2} 
\sim 10^{-3} \gg \Delta.
\eea
Here $\lambda_c =1/m$ is the Compton wavelength, which is $3.86\times 10^{-11}$ for electron, $\sigma_{\perp}$ is the packet width, which is usually larger than $1$ nm for electron  \cite{PhysRevLett.114.227601, cho2004quantitative, cho2013electron,latychevskaia2017spatial,karlovets2021vortex}, and the biggest OAM obtained so far is $|\ell| \sim 1000$ \cite{mafakheri2017realization}.

A non-relativistic Hamiltonian of a charged particle with spin interacting with a constant and homogeneous magnetic field ${\bm H}'=(0,0,H')$ is \cite{Landau1981Quantum}
\bea
\displaystyle & \hat{H} = \fr{\hat{\bp}^2}{2m} -\hat{{\bm \mu}}_s\cdot{\bm H}'=\fr{\left(\hat{\bp}^{(c)}\right)^2}{2m} + \fr{m}{2}\omega_L^2\rho^2 - \hat{{\bm \mu}}^{(c)}\cdot{\bm H}' = \cr\displaystyle & 
= \fr{\left(\hat{\bp}^{(c)}\right)^2}{2m} - \fr{m}{2}\omega_L^2\rho^2 - \hat{{\bm \mu}}\cdot{\bm H}'.
\eea
 Here $\omega_L = eH'/2m$ and ${\bm A} = [{\bm H}'\times \br]/2$. 
The $z$-component of the canonical OAM commutes with the Hamiltonian $[\hat{H},\hat{L}_z^{(c)}]=0,$ which is why $\la\hat{L}_z^{(c)}\ra$ is an integral of motion. On the contrary, the $z$-component of the kinetic OAM {\it oscillates} with time with the cyclotron frequency $\omega_c=eH'/m$ together with the rms radius of the electron packet \cite{karlovets2021vortex, greenshields2014angular, greenshields2015parallel}
\bea
\la\hat{L}_z\ra(\tau) = \la\hat{L}_z^{(c)}\ra + \fr{eH'}{2}\la\rho^2\ra(\tau) \cr
= \text{const} + \alpha_1 \cos\left(\omega_c\tau+\alpha_2\right),
\label{Lzkin}
\eea
where $\tau$ is time in the rest frame, 
and $\alpha_1, \alpha_2$ are constants defined by initial conditions. 

Let us now put $\hat{{\bm \mu}}^{(c)}$ and $\hat{{\bm \mu}}$ in the Heisenberg equations \cite{Landau1981Quantum, mane2005siberian, mane2005spin, silenko2018relativistic}  $d\hat{\bm \mu} / d\tau = \partial\hat{\bm \mu} / \partial\tau+i[\hat{H},\hat{{\bm \mu}}]$.
We suppose that $\partial\hat{{\bm s}}/\partial\tau = \partial\hat{{\bm L}}^{(c)}/\partial\tau=0$, thereby neglecting spin depolarization and OAM losses due to the photon emission. Within the quasi-classical approximation (see End Matter Appendix 1), we find for the mean values 
\bea
\fr{d\la\hat{\bm L}+g\hat{\bm s}\ra}{d\tau} \simeq \mu_B\,{\bm H}'\times\la\hat{\bm L}+g^2\hat{\bm s}\ra,
\label{BMTrestmean}
\eea
and this equation stays the same for the canonical OAM, $\la\hat{\bm L}^{(c)}\ra$, within this approximation. 
Clearly, this equation does {\it not} describe pure precession due to $g^2$ in the r.h.s. Without spin, Eq.(\ref{BMTrestmean}) coincides with Eq.(2.29) in Ref.\cite{Bliokh2017} supporting the conclusion that the orbital g-factor equals 1. When the OAM is omitted, we reproduce the BMT equation for spin in the rest frame \cite{bargmann1959precession, BLP}.



Now we seek a covariant generalization of Eq.(\ref{BMTrestmean}) following the method of sec.41 in \cite{BLP} within the same quasi-classical approximation. We define a space-like vector $a_L$ that in the rest frame reduces to $a_L \to (0,\la\hat{\bm L}\ra)$, analogously to the spin 4-vector $a_s$ \cite{BLP}. Then we define two 4-vectors $a_g = a_L + g\,a_s/2$ and $a_{g^2} = a_L+g^2\,a_s/2, (a_g u) = (a_{g^2}u) =0$, that in the rest frame look as follows:
\bea
\displaystyle & a_g^{\mu} = a_L^{\mu}+\fr{g}{2}a_s^{\mu} \to (0, \la{\bm L}\ra + g\la{\bm s}\ra),\cr
\displaystyle & a_{g^2}^{\mu} = a_L^{\mu}+\fr{g^2}{2}a_s^{\mu} \to (0, \la{\bm L}\ra + g^2\la{\bm s}\ra),
\eea
and $u^{\mu} \equiv \la u^{\mu}\ra$ is a mean velocity of the packet. The result is (see details in the End Matter Appendix 2)
\bea
\fr{da_g^{\mu}}{d\tau} &=& \fr{e}{2m}F^{\mu\nu}{a_{g^2}}_{\nu}\cr
&+& \fr{e}{2m} \left(2 - \fr{F^{\sigma\lambda}u_{\sigma}{a_{g^2}}_{\lambda}}{F^{\sigma\lambda}u_{\sigma}{a_{g}}_{\lambda}}\right) \underbrace{u^{\mu} F^{\nu\eta}u_{\nu}{a_g}_{\eta}}_{\text{Thomas precession}}.
\label{BMT}
\eea
Clearly, in the rest frame with $u'=(1,{\bm 0})$ we return to Eq.(\ref{BMTrestmean}). When $\la\hat{\bm L}\ra=0$, the BMT equation for spin \cite{bargmann1959precession} also follows from Eq.(\ref{BMT}),
\bea
\displaystyle & \fr{d a_s^{\mu}}{d\tau} = \underbrace{\fr{e}{2m}g}_{2\mu_s}F^{\mu\nu}{a_s}_{\nu} - \underbrace{\fr{e}{2m}(g-2)}_{2\mu'} u^{\mu} F^{\nu\eta}u_{\nu}{a_s}_{\eta},
\label{BMTS}
\eea
and the Thomas precession term is due to the AMM $\mu'$. 

In the opposite case of a spin-unpolarized particle with $\la\hat{\bm s}\ra=0$, we arrive at the following relativistic precession equation of the intrinsic OAM:
\bea
\displaystyle & \fr{d a_L^{\mu}}{d\tau} = \fr{e}{2m}F^{\mu\nu}{a_L}_{\nu} + \underbrace{\fr{e}{2m} u^{\mu} F^{\nu\eta}u_{\nu}{a_L}_{\eta}}_{\text{no longer small}}.
\label{BMTL}
\eea
It can be obtained from the BMT equation (\ref{BMTS}) via the following changes:
\bea
\displaystyle & 2\mu_s = \fr{e}{2m}g \to \fr{e}{2m} = -\mu_B,\cr
\displaystyle & 2\mu' = \fr{e}{2m}(g-2) \to - \fr{e}{2m}=\mu_B,
\eea
where the former simply says that the orbital g-factor equals 1. The latter change has a dramatic effect on the precession dynamics because now the second term in the r.h.s. of Eq.(\ref{BMTL}) has a different sign compared to Eq.(\ref{BMTS}) and its magnitude is $1/(g-2) \approx 431$ {\it times larger} than the AMM term in Eq.(\ref{BMTS}). 
As $F^{\nu\eta}u_{\nu} \propto du^{\eta}/d\tau$, the second term in the r.h.s. of Eq.(\ref{BMTL}) is due to the Fermi–Walker transport and the Thomas precession. This term in the BMT equation (\ref{BMTS}) leads to spin resonances starting at the energy of 440 MeV \cite{mane2005siberian, mane2005spin}, defined by the AMM. Importantly, the contribution of this term for vortex particles is no longer small and it shifts the analogous OAM resonances to much lower energies. 


Similar to the BMT equation \cite{mane2005siberian}, we rewrite (\ref{BMTL}) when the intrinsic OAM $\la{\bm L}\ra$ is given \textit{in the rest frame}, whereas all other quantities, including the field strengths ${\bm E}, {\bm H}$, are in the lab frame,
\bea
\displaystyle & \fr{d\la{\bm L}\ra}{dt} = {\bm \Omega}_L \times \la{\bm L}\ra,\cr
\displaystyle & {\bm \Omega}_L = -\fr{e}{2m}\Bigg(\fr{2-\gamma}{\gamma}{\bm H} + \fr{\gamma}{1+\gamma}{\bm u}({\bm u}\cdot {\bm H}) \cr
\displaystyle & + \fr{\gamma-1}{\gamma+1} [{\bm u}\times{\bm E}]\Bigg),\ \gamma=\varepsilon/m.
\label{precL}
\eea
This contrasts with Eq.(10) in \cite{silenko2017manipulating} or Eq.(1) in \cite{silenko2019siberian} where $\la{\bm L}\ra$ is in the lab frame, which can be less convenient for experiments (see the End Matter Appendix 2). It is useful to compare Eq.(\ref{precL}) with (\ref{BMTS}) for spin alone
\bea
& \fr{d\la{\bm s}\ra}{dt} = {\bm \Omega}_s \times \la{\bm s}\ra,\cr
& {\bm \Omega}_s = -\fr{e}{m}\Bigg(\left(a + \fr{1}{\gamma}\right){\bm H} -a\fr{\gamma}{1+\gamma}{\bm u}({\bm u}\cdot {\bm H}) \cr
& -\left(a + \fr{1}{\gamma+1}\right) [{\bm u}\times{\bm E}]\Bigg)
\label{precS}
\eea
where the spin $\la{\bm s}\ra$ is also in the rest frame. The very first observation is that the longitudinal electric field ${\bm E}\uparrow\uparrow{\bm u}$ -- say, in an RF cavity -- {\it influences neither spin nor OAM}. As a result, it is possible to accelerate vortex particles in linacs while preserving the beam OAM if one can neglect the OAM radiative depolarization. 

Let us discuss now the role of magnetic fields during the acceleration. In a solenoid with ${\bm H}\uparrow\uparrow{\bm u}$, we have
\bea
\displaystyle & {\bm \Omega}_{L,||} = -\fr{e}{2m} \fr{1}{\gamma}{\bm H},\quad {\bm \Omega}_{s,||} = -\fr{e}{m}\fr{a+1}{\gamma}{\bm H},\cr
\displaystyle & \fr{|{\bm \Omega}_{L,||}|}{{|\bm \Omega}_{s,||}|} = \fr{1}{2(a+1)} \approx 1/2.
\label{precLL}
\eea
In the {\it transverse} magnetic field of a solenoid fringe field or a quadrupole magnet, we have ${\bm u}\cdot {\bm H}=0$ and
\bea
\displaystyle & {\bm \Omega}_{L,\perp} = -\fr{e}{2m}\fr{2-\gamma}{\gamma}{\bm H},\quad {\bm \Omega}_{s,\perp} = -\fr{e}{m}\left(a + \fr{1}{\gamma}\right){\bm H},\cr
\displaystyle & \fr{|{\bm \Omega}_{L,\perp}|}{{|\bm \Omega}_{s,\perp}|} = \fr{1}{2}\,\fr{|\gamma-2|}{a\gamma+1}.
\label{precLL}
\eea
This formula for ${\bm \Omega}_{L,\perp}$ is also applicable when the OAM represents {\it the orbital helicity} \cite{ivanov2022promises}, $\la{\bm L}\ra \uparrow\uparrow {\bm u}$, so the longitudinal component of the magnetic field does {\it not} contribute to ${\bm \Omega}_{L,\perp}$. 

Clearly, there is a ''magic`` energy $\gamma=2$ where the OAM precession stops, ${\bm \Omega}_{L,\perp}\to 0$, and ${\bm \Omega}_{L,\perp}$ has different signs for $\gamma<2$ and $\gamma>2$, so ${\bm \Omega}_{L,\perp} \uparrow\downarrow {\bm \Omega}_{s,\perp}$ for $\gamma>2$. There are following three regimes with the distinct {\it orbit-to-spin ratios}:
\bea
&1 < \gamma <2:& \fr{|{\bm \Omega}_{L,\perp}|}{{|\bm \Omega}_{s,\perp}|} = |2-\gamma|/2 \in [0,1/2],\cr
&2 \ll \gamma \ll a^{-1}\sim 10^3:& \fr{|{\bm \Omega}_{L,\perp}|}{{|\bm \Omega}_{s,\perp}|} \approx \gamma/2 \gg 1,\cr
& a^{-1}\sim 10^3 \ll \gamma:& \fr{|{\bm \Omega}_{L,\perp}|}{{|\bm \Omega}_{s,\perp}|} \approx 1/2a \approx 430.
\label{ratiotr}
\eea
The spin precesses faster than OAM in the former regime, whereas the OAM precesses faster than spin in the two later ones,
and the value $1/2a \approx 430$ represents the upper limit of the ratio. For vortex electrons with $\varepsilon \sim 5 - 200\,\text{MeV}$, the scaling $\gamma/2$ can be used for determining the beam energy, whereas for GeV-range electrons one can determine the AMM by measuring the orbit-to-spin ratio.
Note that in a Wien filter with the orthogonal fields and the vanishing Lorentz force ${\bm E} + {\bm \beta}\times{\bm H}=0$, we have
\bea
\displaystyle & {\bm \Omega}_{L} = -\fr{e}{2m}\fr{1}{\gamma^2}{\bm H},\ {\bm \Omega}_{s} = -\fr{e}{2m}\fr{g}{\gamma^2}{\bm H},\cr \displaystyle & |{\bm \Omega}_{L}|/{|\bm \Omega}_{s}|=1/g.
\eea

The same formulas (\ref{precLL}), (\ref{ratiotr}) hold in the transverse magnetic field of a storage ring or a synchrotron with ${\bm u} \perp {\bm H}, {\bm E}=0$, when the packet centroid revolves with the frequency $\omega_c = |e|H/\gamma m$ in the lab frame.
The spin tune is defined as \cite{mane2005spin}
\bea
\displaystyle & \nu_s = \fr{|{\bm \Omega}_{s,\perp}|-\omega_c}{\omega_c} = a\gamma, 
\eea
so the first spin-depolarizing resonance occurs at $\nu_s = a\gamma = 1$ \cite{mane2005spin, mane2005siberian}, $\varepsilon = 2m/(g-2) \approx 440.65\,\text{MeV}$.

For OAM precession, we find {\it the OAM tune}
\bea
\displaystyle & \nu_L = \fr{|{\bm \Omega}_{L,\perp}|-\omega_c}{\omega_c} = \fr{|2-\gamma|}{2}-1, 
\eea
and there are two distinct regimes:
\bea
\displaystyle && 1<\gamma<2:\ \nu_L = -\gamma/2 \in [-1, -1/2],\cr
\displaystyle && \gamma>2:\ \nu_L = \gamma/2-2.
\eea
Analogously to spin depolarizing resonances \cite{mane2005spin, mane2005siberian}, there can be ``OAM-depolarizing'' resonances that can {\it non-radiatively} destroy the beam OAM at $\nu_L = k = \pm 1,\pm 2,\pm 3,...$. For $1<\gamma<2$ there are no resonances because the precession stops at $\gamma=2$, but for $\gamma>2$ there are many resonances with the step $\Delta \gamma=2$ or $\Delta \varepsilon = 1.022$ MeV for electron:
\bea
\displaystyle & \nu_L = \gamma/2-2=k,\ \gamma_{\text{res}}^{(k)} = 2(k+2) = 6,8,10,...,
\label{OAMres}
\eea
the first resonance being at the energy of 
\bea
\varepsilon = m\gamma_{\text{res}}^{(1)} \approx 3.066\, \text{MeV},
\eea
two orders of magnitude less than 440 MeV for spin. This is because the contribution of the second (Thomas precession) term in the r.h.s of Eq.(\ref{BMTL}) is much stronger than that of the AMM term in the BMT equation (\ref{BMTS}).

\begin{figure*}[t!]
	\centering
	\begin{subfigure}{0.4\linewidth}
	\includegraphics[width=\linewidth]{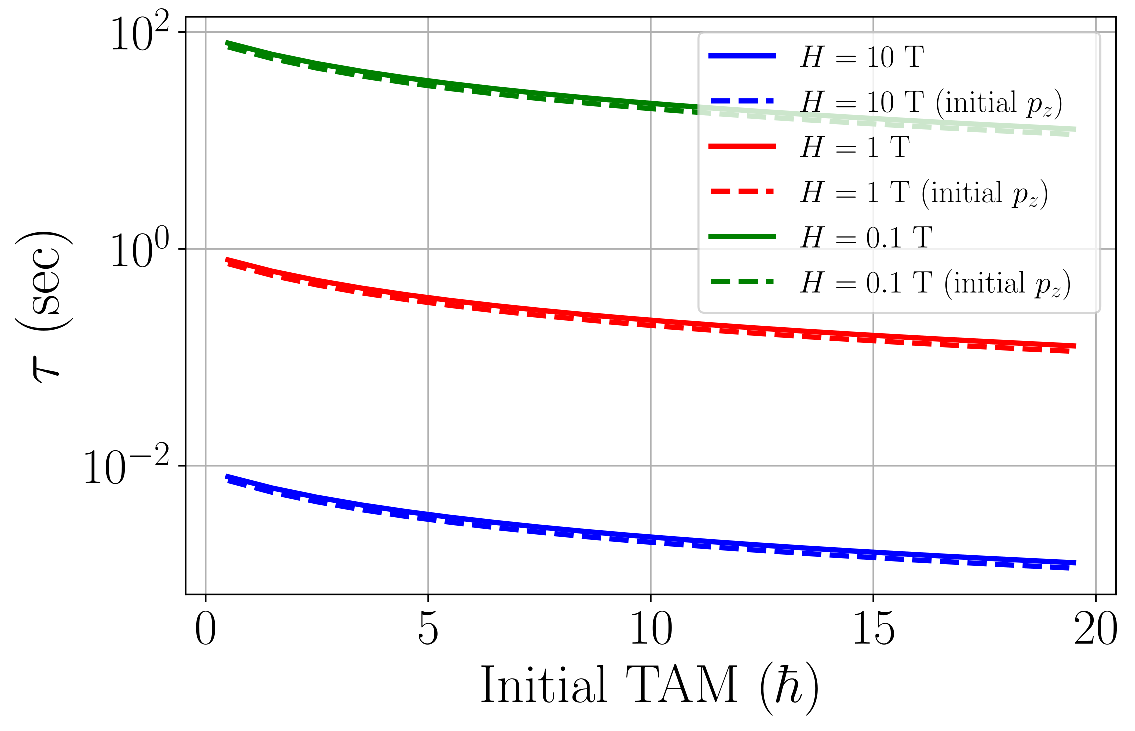}
	\end{subfigure}
	\begin{subfigure}{0.4\linewidth}
		\includegraphics[width=\linewidth]{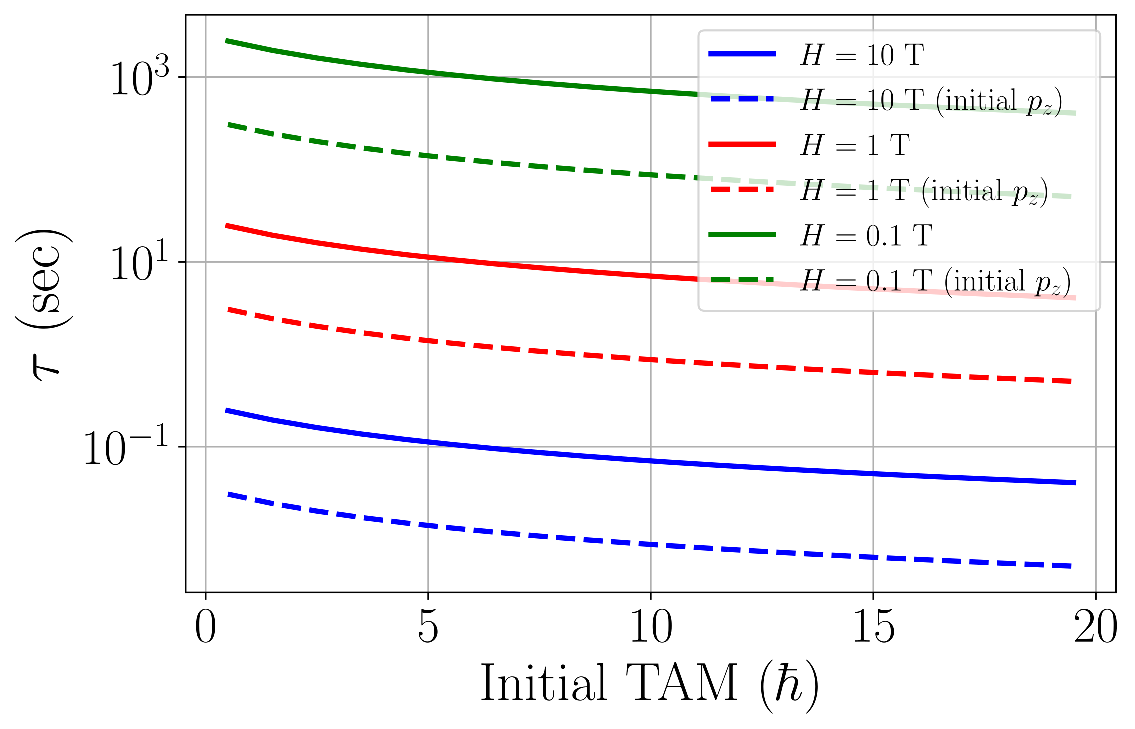}
	\end{subfigure}
	\caption{Effective lifetime of vortex electron states with the radial quantum number $s=3$. The solid lines show the lifetime after accounting for electron acceleration by averaging the emission probability over the longitudinal momenta $p_z$, while dashed lines correspond to the states with a fixed initial momentum (no acceleration). The averaging spans kinetic energies of $1$ keV -- $1$ MeV (left) and $5$ -- $200$ MeV (right)}
 \label{e_lifetime}
\end{figure*}

Thus acceleration of spin-unpolarized vortex particles in a ring can be {\it much less effective} than that of spin-polarized beams without OAM due to the much smaller step between depolarizing resonances, $1.022$ MeV instead of $440.65$ MeV for electrons. 
This drawback can be overcome with the help of Siberian snakes or spin rotators \cite{mane2005siberian, mane2005spin, silenko2019siberian} rotating the OAM instead of spin. The quantum states with the OAM directed {\it not} along the mean momentum are called the {\it spatiotemporal vortex beams} \cite{Bliokh2012}, and the snakes can be used to make this state an ordinary vortex beam again. Alternatively, one can deliberately prepare the spatiotemporal vortex beams with the needed angle between the mean momentum and the OAM by using the snakes.

To avoid OAM resonances (\ref{OAMres}), one can also utilize {\it spin-polarized} vortex electrons because the spin precesses with a different frequency and it renders the OAM resonance condition imperfect. The opposite is also true: if a spin-polarized particle is converted to a twisted state -- say, with a magnetic or electric needle \cite{beche2014magnetic, tavabi2020generation} -- the OAM can help to pass the spin resonances, after which one can remove the OAM with a needle of an opposite polarity. Interestingly, the quantum terms (see End Matter Appendix 1), omitted in (\ref{BMTrestmean}), (\ref{BMT}), can also help with the passing of resonances.

Acceleration of vortex electrons to a few MeV can be achieved more easily with DC- and photoelectron guns, whereas the GeV range can be reached at linacs with RF cavities. The above model with the homogeneous and stationary fields holds because the longitudinal spatial coherence of a massive packet is many orders of magnitude smaller than a wavelength of the accelerating cavity. 

\begin{figure}[h]
	\centering
	\includegraphics[width=0.8\linewidth]{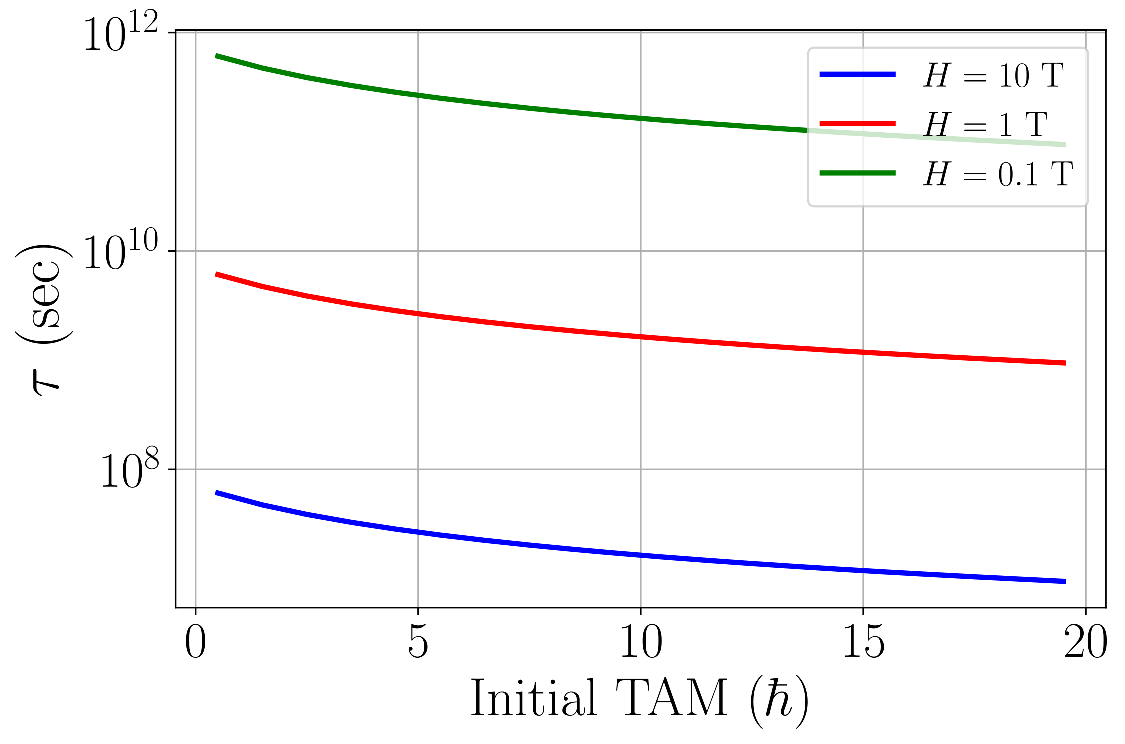}
	\caption{Effective lifetime of the vortex state for a $^{12}\mathrm{C}^{6+}$ ion with the radial quantum number $s=3$, accounting for the acceleration effects by averaging the emission probability over kinetic energies from $10$ keV/nucleon to $10$ MeV/nucleon.}
 \label{ion_lifetime}
\end{figure}

\textit{Radiation loss of the OAM---} When vortex particles are accelerated along the $z$-axis in a linac, the transverse magnetic field can rotate the magnetic moments, whereas the longitudinal field $\bm{H} = (0,0,H)$ can transversely focus the beam and induce transitions between the Landau levels. As a result, electrons can spontaneously go to lower levels by emitting twisted photons, with the ``effective lifetime'' of the twisted state increasing with particle energy due to Lorentz time dilation: at rest, the lifetime is $\gamma$ times shorter than at the energy $\varepsilon = \gamma m$. The transverse rms-size of electron packets in the field is $1$–$100$ nm (see Eq. \eqref{rho_H} in End Matter Appendix), while solenoidal field inhomogeneities manifest at scales $\gtrsim 1~\mu$m. Thus, each packet experiences a locally homogeneous and constant field—--an approximation that improves for relativistic particles.

To calculate the lifetime, we use the first-order QED in the Furry picture, where the states of fermions in magnetic field are described with exact solutions of the Dirac equation \cite{ST1986}. The relativistic Landau states are characterized by three quantum numbers: the radial number $s=0,1,2,\dots$, the half-integer \textit{total} angular momentum (TAM) projection $j_z$ (combining OAM and spin), and the longitudinal momentum $p_z$. An additional spin degree of freedom permits two orthogonal states for fixed $s$, $j_z$, and $p_z$, which are not uniquely defined due to the ambiguity of a relativistic spin operator \cite{bordovitsyn1999synchrotron, bauke2014relativistic, pavlov2024emission}.

The lifetime of an excited Landau state is obtained by computing the transition matrix elements for all kinematically allowed final states. These are given by the standard QED expression $S_{fi} = -ie\int d^4 x\, j^\mu_{fi} A_\mu (x)^*,$
where $A_\mu (x)$ is the photon four-potential and $j^\mu_{fi}$ is the transition current. The transition probability per unit time is derived by integrating $|S_{fi}|^2$ over the complete basis of emitted photon states (see \cite{pavlov2024emission} for details).
\be
    \dot{W}^{(1)}_{s', j_z'}(p_z') = \fr{1}{T} \sum_{\lambda=\pm 1} \int\frac{\dd^3 k}{(2\pi)^3} |S_{fi}|^2,
\ee
where $T$ is the normalization time. The total probability is given by the summation over all possible quantum numbers of the final Landau states, $\dot{W}^{(1)} = \sum_{s', j_z'} \int \dot{W}^{(1)}_{s', j_z'}(p_z') d p'_z/2\pi$,
where the normalization is omitted. We also average the probability over the initial polarizations and sum over the final ones. The effective lifetime of an excited state is defined as $\tau \equiv 1 / \dot{W}^{(1)}$. While the Landau states' infinite degeneracy allows for infinitely many final states, transitions with large $\Delta j_z$ or $\Delta s$ are strongly suppressed \cite{pavlov2024emission}, making a finite subset sufficient in practice. Calculation details appear in \cite{pavlov2024emission,karlovets2023emission}, with the latter paper treating scalar particles. Our current model approximates the effect of electric field acceleration by averaging the emission probability over the initial $p_z$ values. A more complete treatment using quantum states in combined electric and magnetic fields yields \cite{murtazin2025photon} consistent results.

Figures~\ref{e_lifetime} and~\ref{ion_lifetime} present the  lifetimes of vortex states for electrons and carbon ions ($^{12}$C$^{6+}$) in magnetic fields of $0.1-10$ T, far below the critical field $H_c = 4.4 \times 10^9$ T, resulting in the approximate scaling $\tau \propto H^{-2}$. For ions, ``vortex'' refers only to center-of-mass OAM (not internal bound electrons/nucleons). The electron lifetimes range from $10^{-2}$ to $10^3$ seconds --- many orders of magnitude longer than typical linac transit times (e.g. for a 1 km linac, $1\,\text{km}/3\cdot 10^8\,\text{m}/\text{s} \lesssim 10^{-5}$ s ). Carbon ions exhibit even longer lifetimes, $\tau \gtrsim 10^7$ s, due to the approximate $\tau \propto m^3$ mass scaling. Transitions with $\Delta j_z = -1$ dominate \cite{pavlov2024emission,karlovets2023emission}, making the $|j_z| \gg 1 \to j_z=0$ transition lifetime even longer than shown in the figures.

Even with field inhomogeneities (e.g., at solenoid edges) potentially reducing $\tau$ by $10$--$100$ times, lifetimes remain much longer than acceleration times. Although longitudinal quantization during acceleration could be incorporated in the QED formalism, it seems unlikely to significantly reduce $\tau$.

\textit{Summary---} We have studied the quantum dynamics of angular momentum of vortex particles in the fields typical for accelerators. Whereas radiative OAM loss is negligible, non-radiative effects — such as differing OAM and spin precession rates — are critical. In circular accelerators, Siberian snakes can be essential for unpolarized vortex beams due to many OAM-depolarizing resonances, the first of them being at 3 MeV for electrons. At such energies, the compact machines such as microtrons can be employed to probe these resonances, while electron diffraction can be utilized to analyze the OAM \cite{maksimov2025diffraction}. 

Linacs offer a simpler alternative to reach ultrarelativistic energies, whereas Siberian snakes can be used to generate the spatiotemporal vortex beams of electrons, protons, or ions on demand, thereby enabling completely new types of experiments in particle physics. However, the impact of space charge on quantum states remains unknown, particularly for vortex electrons of a few MeV, which potentially limits the luminosity of such accelerators and it warrants further study.\\

\textit{Acknowledgment---}We are grateful to A.\, Silenko, E.\,Akhmedov, G.\,Sizykh, D.\, Naumov, and I.\,Ivanov for fruitful discussions. The studies on the non-radiative dynamics of magnetic moments are supported by the Russian Science Foundation, project No. 25-71-00060. The studies on the quantum corrections to classical precession are supported by the Foundation for the Advancement of Theoretical Physics and Mathematics “BASIS”. The studies on the radiative dynamics in accelerators are supported by the Russian Science Foundation, project No. 23-62-10026 \cite{JINR}.

\bibliographystyle{apsrev}
\bibliography{references}

\clearpage

\appendix
\textit{Appendix 1: Quasi-classical approximation---} Consider a non-relativistic electron with a charge and magnetic moment in the magnetic field $H'$ with the potential ${\bm A} = [{\bm H}'\times \br]/2$. For the canonical magnetic moment operator (\ref{munosoi}), from the Heisenberg equation of motion we find
 \bea
\displaystyle & \fr{d\left(\hat{\bm L}^{(c)}+g\hat{\bm s}\right)}{d\tau} = \mu_B\,{\bm H}'\times\left(\hat{\bm L}+g^2\hat{\bm s}\right),
\label{BMTrest}
\eea
where 
\bea
{\bm H}'\times \hat{\bm L}={\bm H}'\times (\hat{\bm L}^{(c)}+\fr{e}{2}\br({\bm H}'\cdot\br)),
\eea
$\br =\{{\bm \rho}, z\}=\{x, y, z\}$, and we have used standard angular momentum commutation relations and
\bea
\displaystyle & \left[\left(\hat{\bp}^{(c)}\right)^2, \hat{\bm L}^{(c)}\right]=0,\cr
\displaystyle &
[\rho^2, \hat{L}_z^{(c)}] = -2iz(y,-x,0)=4i{\bm A}(\br\cdot {\bm H'})/(H')^2.
\eea

If we plug the kinetic magnetic moment into the Heisenberg equation, the analogous result is
\bea
\displaystyle & \fr{d\left(\hat{\bm L}+g\hat{\bm s}\right)}{d\tau} = \mu_B\,{\bm H}'\times\left(\hat{\bm L}+g^2\hat{\bm s}\right) - \fr{d [\br \times e{\bm A}]}{d\tau}.
\label{BMTrestkin}
\eea
Here
\bea
&-\fr{d [\br \times e{\bm A}]}{d\tau} = -i[\hat{H}, \br \times e{\bm A}]\cr
&=-\mu_B\,H'\,
\begin{pmatrix}
z\hat{p}_x^{(c)}+x\hat{p}_z^{(c)} + \fr{eH'}{2}yz \\
z\hat{p}_y^{(c)}+y\hat{p}_z^{(c)} - \fr{eH'}{2}xz\\
-2(-i+x\hat{p}_x^{(c)}+y\hat{p}_y^{(c)})
\end{pmatrix},
\eea
and one can use
$d\br/d\tau = \hat{{\bm p}}/m=\left(\hat{{\bm p}}^{(c)}-e{\bm A}\right)/m$.

The OAM dynamics along the field ${\bm H}'$ is defined by Eq.(\ref{Lzkin}), so we take a closer look at $x-$ and $y-$ components of Eq.(\ref{BMTrest}) and Eq.(\ref{BMTrestkin}). Clearly, neither of these equations describes pure precession because the length of the vector $\la \hat{{\bm \mu}}\ra \propto \la \hat{\bm L}+g\hat{\bm s}\ra$ also changes in time due to $g^2$ in the r.h.s. So the OAM $\la \hat{\bm L}\ra$ and the spin $\la\hat{\bm s}\ra$ precess independently with different angular velocities due to $g=1$ for orbital motion. 

Moreover, there are quantum dynamic terms due to $[\rho^2, \hat{L}_z^{(c)}] \ne 0$. When averaged over some quantum state of the electron in magnetic field, these quantum terms are proportional to
\bea
&\fr{e}{2}\la{\bm \rho}({\bm H}'\cdot\br)\ra \propto  eH' \left(\la xz\ra,\la yz\ra\right),\cr
 &\left\la\fr{d [\br \times e{\bm A}]_{\perp}}{d\tau}\right\ra \propto 
\begin{pmatrix}
\la z\hat{p}_x^{(c)}\ra+\la x\hat{p}_z^{(c)}\ra + \fr{eH'}{2}\la yz\ra\\
\la z\hat{p}_y^{(c)}\ra+\la y\hat{p}_z^{(c)}\ra - \fr{eH'}{2}\la xz\ra \end{pmatrix}
.
\eea
In many practical cases, there are {\it no quantum correlations} between the motion along the field and in the perpendicular plane, and so 
$\la xz\ra = \la x\ra\la z\ra=0, \la z\hat{p}_x^{(c)}\ra=\la z\ra\la\hat{p}_x^{(c)}\ra=0$, etc. This is so, for instance, when a particle moves along the central axis of a linac with no deflecting fields like those of quadrupoles and therefore $\la x\ra=\la y\ra =0$ or in a storage ring in a reference plane with $\la z\ra=0$. 

In reality, these quantum terms are {\it not vanishing}, but their contribution is very small for the fields typical for accelerators. Indeed, the contribution of, say, $|e|H'\la yz\ra/2$
needs to be compared with $\la {\hat L_x}^{(c)}\ra$, which is why it becomes important only when 
\bea
|e|H'2\la yz\ra/2 = 2\la yz\ra/\rho_H^2 \gtrsim 1\, (\equiv \hbar).
\eea
Here
\bea
\label{rho_H}
\rho_H = 2/\sqrt{|e|H'} = 2\lambda_c\sqrt{H_c/H'}
\eea
is a radius of a ground Landau state and $H_c = 4.4\times 10^9$ T is the Schwinger critical field \cite{ST1986}. For the fields in the electron rest frame $H'\sim 1-100$ T, we find $\rho_H/\sqrt{2} \sim 3.6-36\, \text{nm}$,
whereas the transverse rms sizes of non-relativistic electron packets at room temperature, $\sqrt{\la x^2\ra}, \sqrt{\la y^2\ra}$, or $\sqrt{\la \rho^2\ra}$, {\it do not exceed a few nanometers} for standard photo-, DC- or thermal guns \cite{PhysRevLett.114.227601, cho2004quantitative, cho2013electron,latychevskaia2017spatial,karlovets2021vortex}. Therefore, the correlations like $\la xz\ra, \la z\hat{p}_x^{(c)}\ra$ can hardly survive at the distances much larger than that, and so
\bea
|e|H'\la yz\ra/2 = 2\la yz\ra/\rho_H^2 < 1\ \text{for}\ H' \lesssim 10\, \text{T}.
\eea
Importantly, these quantum terms can become important for the magnetic fields much stronger than 10 T -- say, those of the neutron stars where the field can reach the critical value in the lab frame, $H \sim H_c$.

Omitting the above quantum dynamic terms is a part of {\it the quasi-classical approximation} in which a point-like particle moves along a classical path, usually implied when deriving the BMT equation for spin precession \cite{bargmann1959precession, mane2005siberian, BLP}. Similarly, we disregard contributions from higher-order Feynman diagrams and gravitational fields. Although electron self-interaction has been shown to influence spin dynamics under strong-field conditions(\cite{meuren2011quantum, li2023strong}), its impact is expected to be insignificant for the weak fields typical for accelerators. By extension, we posit that this insignificance applies equally to OAM dynamics. Curiously, when $\la\hat{\bm L}\ra \to 0$ but the above dynamic terms are not neglected, the BMT equation can acquire quantum corrections that can become important in the strong fields.

\

\textit{Appendix 2: The relativistic precession equation}

Following the method of sec.41 in \cite{BLP}, we can derive a covariant generalization of equation of precession (\ref{BMTrestmean}). The classical total angular momentum tensor consists of the orbital part $L^{\mu\nu} = x^{\mu}p^{\nu} - x^{\nu}p^{\mu} = (\br-{\bm u}t, \br \times \bp)$ and the spin part $S^{\mu\nu}$. The former is time-like because in the rest frame $L^{\mu\nu} \to (\br, {\bm 0})$, while the latter is space-like.  In quantum theory, we distinguish an {\it extrinsic} OAM tensor, 
$ L^{\mu\nu}_{\text{ext}} = \la \hat{x}\ra^{\mu}\la \hat{p}\ra^{\nu} - \la \hat{x}\ra^{\nu}\la \hat{p}\ra^{\mu}$,
which is also time-like, and its {\it intrinsic} counterpart
\bea
& \la \hat{L}^{\mu\nu}\ra_{\text{int}} = \la \hat{x}^{\mu} \hat{p}^{\nu}\ra - \la \hat{x}^{\nu} \hat{p}^{\mu}\ra = \left(\la \hat{\br}\ra - \la \hat{{\bm u}}\ra t, \la \hat{\br} \times \hat{\bp}\ra\right),
\eea
where the brackets denote quantum averaging and ${\bp} \equiv {\bp}^{(c)} = -i {\bm \nabla}$. In the rest frame of the packet, the mean momentum vanishes and its centroid rests in the origin, so {\it this tensor is space-like}, $\la \hat{L}^{\mu\nu}\ra_{\text{int}} \xrightarrow[]{\text{rest frame}} ({\bm 0}, \la \hat{\br} \times \hat{\bp}\ra)$.

One can avoid the apparent difficulty that the extrinsic and intrinsic OAM tensors transform differently by considering the tensor of intrinsic electric and magnetic dipole moments. Since the intrinsic electric dipole moment vanishes, this tensor is spacelike; in the packet rest frame it reads $({\bm 0}, \langle \hat{\bm \mu} \rangle)$, where $\langle \hat{\bm \mu} \rangle = -\mu_B \langle \hat{\bm L} + g \hat{\bm s} \rangle$ [Eq.~(\ref{mukin})]. Hence, the tensor is dual to a spacelike four-vector. Although this tensor can be Lorentz-transformed to the laboratory frame \cite{silenko2017manipulating}, we instead adopt an alternative approach by decomposing the magnetic moment into orbital and spin contributions (neglecting SOI).

The orbital part of the magnetic moment is described with a space-like 4-vector $a_L$ that in the rest frame reduces to $a_L \xrightarrow[]{\text{rest frame}} (0,\la\hat{\bm L}\ra)$,
and so $\la u^{\mu}\ra (a_L)_{\mu}=0$. Here $\la\hat{\bm L}\ra$ is intrinsic because the origin of the rest frame coincides with the packet's centroid (recall that ${\bm r}_0=0$) and within the above quasiclassical approximation $\la\hat{\bm L}\ra$ can be either canonical or kinetic. Similar to the analogous spin vector $a_s \to (0, 2\la\hat{\bm s}\ra)$ \cite{BLP}, the vector $a_L$ is dual to $\la L^{\mu\nu}\ra_{\text{int}}$, so that 
\bea
& a_L^{\mu} = -2\epsilon^{\mu\nu\eta\rho}\la\hat{ L}_{\nu\eta}\ra_{\text{int}} \la \hat{u}_{\rho}\ra,\cr 
& \la \hat{L}^{\mu\nu}\ra_{\text{int}} = \epsilon^{\mu\nu\eta\rho}(a_L)_{\eta}\la \hat{u}_{\rho}\ra/2, 
\eea
where $\la \hat{u}^{\mu}\ra=\la\hat{ p}^{\mu}\ra/m = d\la \hat{r}^{\mu}\ra/d\tau$ is a mean 4-velocity. From now on, we write $u^{\mu}$ instead of $\la \hat{u}^{\mu}\ra$ for brevity. 

Keeping in mind that $g=\text{inv}$, let us define two 4-vectors $a_g = a_L + g\,a_s/2$ and $a_{g^2} = a_L+g^2\,a_s/2$ that in the rest frame look as follows:
\bea
\displaystyle & a_g^{\mu} = a_L^{\mu}+\fr{g}{2}a_s^{\mu} \xrightarrow[]{\text{rest frame}} (0, \la{\bm L}\ra + g\la{\bm s}\ra),\cr
\displaystyle & a_{g^2}^{\mu} = a_L^{\mu}+\fr{g^2}{2}a_s^{\mu} \xrightarrow[]{\text{rest frame}} (0, \la{\bm L}\ra + g^2\la{\bm s}\ra),
\eea
$(a_g u) = (a_{g^2}u) = 0$. Now we write a general ansatz for a relativistic generalization of Eq.(\ref{BMTrestmean}):
\bea
\displaystyle & \fr{d a_g^{\mu}}{d\tau} = \alpha\, F^{\mu\nu}(a_L)_{\nu} + \beta\, F^{\mu\nu}(a_s)_{\nu} + \cr
\displaystyle & + \gamma\, u^{\mu} F^{\nu\eta}u_{\nu}(a_L)_{\eta} + \delta\, u^{\mu} F^{\nu\eta}u_{\nu}(a_s)_{\eta},
\label{ansatz}
\eea
where $F^{\mu\nu} = \partial^{\mu}A^{\nu}-\partial^{\nu}A^{\mu}$, and $\alpha, \beta, \gamma, \delta$ are the factors to be found. We demand that this equation reduce to Eq.(\ref{BMTrestmean}) in the rest frame and find
\bea
\alpha = \gamma = \fr{e}{2m},\ \beta = \fr{e}{2m} \fr{g^2}{2},\ \delta = \fr{e}{2m} g (1-g/2),
\eea
so the final result is
\bea
\fr{da_g^{\mu}}{d\tau} &=& \fr{e}{2m}F^{\mu\nu}{a_{g^2}}_{\nu}\cr
&+& \underbrace{\fr{e}{2m} 2 u^{\mu} F^{\nu\eta}u_{\nu}{a_g}_{\eta} - \fr{e}{2m} u^{\mu} F^{\sigma\lambda}u_{\sigma}{a_{g^2}}_{\lambda}}_{\text{Thomas precession}}.
\eea

The terms $u^{\mu} F^{\nu\eta} u_{\nu} {a_L}{\eta}$ and $u^{\mu} F^{\nu\eta} u{\nu} {a_s}_{\eta}$ vanish in the rest frame but must appear in the general ansatz (\ref{ansatz}) and are associated with Fermi–Walker transport and Thomas precession. When the magnetic moment $\langle \hat{\bm \mu} \rangle$ is transformed to the laboratory frame, it can undergo Thomas precession purely due to relativistic kinematics. Since it is the magnetic rather than the mechanical moment that precesses in electromagnetic fields, and both OAM and spin contribute to it, these terms affect not only spin dynamics but also OAM precession. 

The Thomas precession terms were not included in the corresponding OAM dynamics equation of Ref.\cite{silenko2017manipulating} due to a different physical model of a ``charged cloud'' that the authors followed. Namely, they first transformed the magnetic moment to the lab frame, then applied the Poisson brackets, and the final equation described the dynamics of the OAM vector in the lab frame without Thomas precession.

Whether the OAM-induced magnetic moment of vortex particles undergoes Thomas precession can be tested experimentally via the OAM-depolarizing resonances discussed in the main text. If the OAM polarization of electrons in a circular accelerator rapidly degrades near 3 MeV, analogous to spin depolarization at 440 MeV, this would directly indicate the presence of Thomas precession. The OAM of 3 MeV vortex electrons could, in principle, be probed using the diffraction-based methods proposed in Ref.~\cite{maksimov2025diffraction} in a drift section of the accelerator.

To put it in more detail, let us compare the angular velocity in the laboratory frame ${\bm \Omega}_L$ derived above with that in Eq.(10) of Ref.\cite{silenko2017manipulating} where we first transform their OAM vector to the rest frame. Note that Ref.\cite{silenko2018relativistic} states that the equations of their more advanced quantum model agree with those in Ref.\cite{silenko2017manipulating}. When $\la {\bm L}\ra \parallel {\bm u}$, which is so for the vortex packets, the OAM in the lab frame simply coincides with that in the rest frame, $\la {\bm L}\ra = \la {\bm L}\ra_{\text{lab}}$.
If no electric field is applied, ${\bm E}=0$, the particle energy is conserved and the angular velocities in the longitudinal magnetic field, ${\bm H} \parallel {\bm u}$, {\it coincide}:
\bea
& {\bm \Omega}_L^{\text{our eq.(\ref{precL})}} = {\bm \Omega}_L^{\text{eq.}(10) \text{from \cite{silenko2017manipulating}}} = -\fr{e}{2m\gamma}{\bm H}.
\eea
For the transverse magnetic field, ${\bm H} \perp {\bm u}$, we find
\bea
\fr{{\bm \Omega}_L^{\text{our eq.}(\ref{precL})}}{{\bm \Omega}_L^{\text{eq.}(10) \text{from \cite{silenko2017manipulating}}}} = 2-\gamma \leq 1.
\eea
The latter ratio turns to unity in the rest frame with $\gamma=1$, but the OAM precession in our model \textit{stops} at $\gamma=2$, and the ratio becomes roughly $-\gamma$ for $\gamma\gg 1$. This big difference is due to the Thomas precession, and its existence for the OAM can be tested experimentally by studying the above OAM-depolarizing resonances in circular accelerators, the first of them being at $\gamma_{\text{res}}^{(1)} = 6$. Even compact accelerators such as \textit{microtrons in the few-MeV range} are suitable for these purposes.

\end{document}